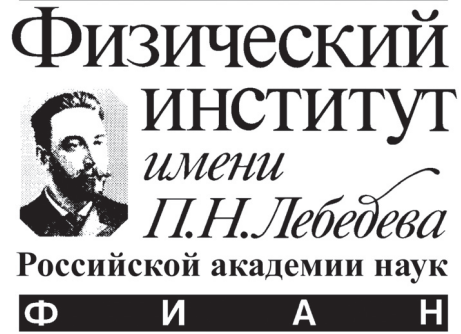

ПРЕПРИНТ

**8**

O.D. DALKAROV, M.A. NEGODAEV, A.S. RUSETSKII

# INVESTIGATION OF HEAT RELEASE IN THE TARGETS DURING IRRADIATION BY ION BEAMS

## Information

O.D. Dalkarov, M.A. Negodaev, A.S. Rusetskii

# INVESTIGATION OF HEAT RELEASE IN THE TARGETS DURING IRRADIATION BY ION BEAMS

№ 8

# Investigation of heat release in the targets during irradiation by ion beams

O.D. Dalkarov, M.A. Negodaev, A.S. Rusetskii*

*rusets@lebedev.ru

*Lebedev Physical Institute (LPI) RAS, Moscow, Russia*

ABSTRACT

The DD-reaction is investigated and the heat emission off the targets during their irradiation with ion beams is studied at the HELIS ion accelerator at LPI. The heat emission is observed to be significantly higher in the case of irradiation of the Ti/TiO$_2$:D$_x$-targets by a D+ beam, as compared to the H+ and Ne+ beams. Furthermore, it depends on the concentration of deuterium in the target and current density of the deuteron beam.

In the earlier experiments at the HELIS facility at LPI, the yields of DD-reactions at energies of 10 - 25 keV have been investigated. It has been shown, that the potential of screening for structures Pd/PdO:D$_x$ and Ti/TiO$_2$:D$_x$ strongly depends on the external conditions like the beam current and the target temperature [1,2]. Indications to a possible stimulation of DD-fusion reactions in the structures of Pd/PdO:D$_x$ and Ti/TiO$_2$:D$_x$ are observed, once irradiated with proton and heavy ion (Ne +) beams [3-5].

For a more precise determination of the yield in DD-reaction, it is important to control the surface temperature of the irradiated target, which affects the effective concentration of deuterium. The experimental setup is shown schematically in Fig. 1a. The investigated target was placed in a water-cooled holder. The diameter of the ion beam passing through the aperture was 6 mm. The surface temperature of the irradiated target was measured by the thermocouple thermometer, based on the chromel-alumel alloy. The energy and the current of the beam, the temperature of the target and of the cooled substrate were monitored during the experiment. DD-reaction neutron yield along and across beam direction was measured by 2 $^3$He-counter groups of neutron detector (Fig.1b)

The Ti/TiO$_2$:D$_x$ targets are made of Ti foil with a thickness of 55 microns and are deuterated by electrolysis in a 0.2M solution of D$_2$SO$_4$ in D$_2$O. The details of this procedure can be found in [2].

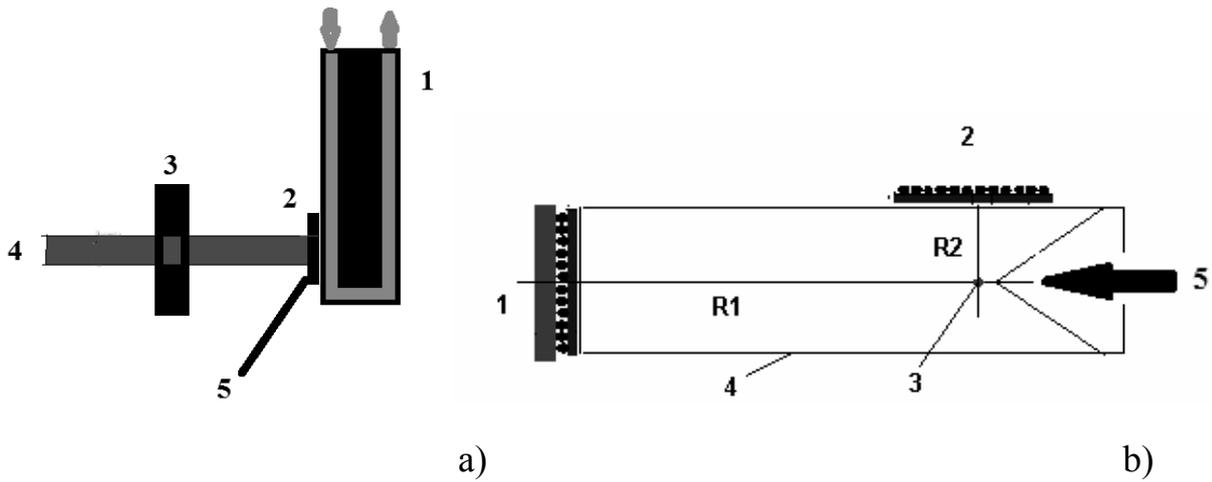

Fig.1. (a) The experimental scheme. (1) a water-cooled target holder; (2) the target; (3) the aperture; (4) the ion beam; (5) the thermocouple thermometer. (b) The $^3$He detector setup at HELIS, representing the first (1) and the second (2) $^3$He-counter groups with radii R=85 cm and R=38 cm, respectively. The target is placed at (3) inside the HELIS beam pipe (4). The ion beam direction is indicated by (5).

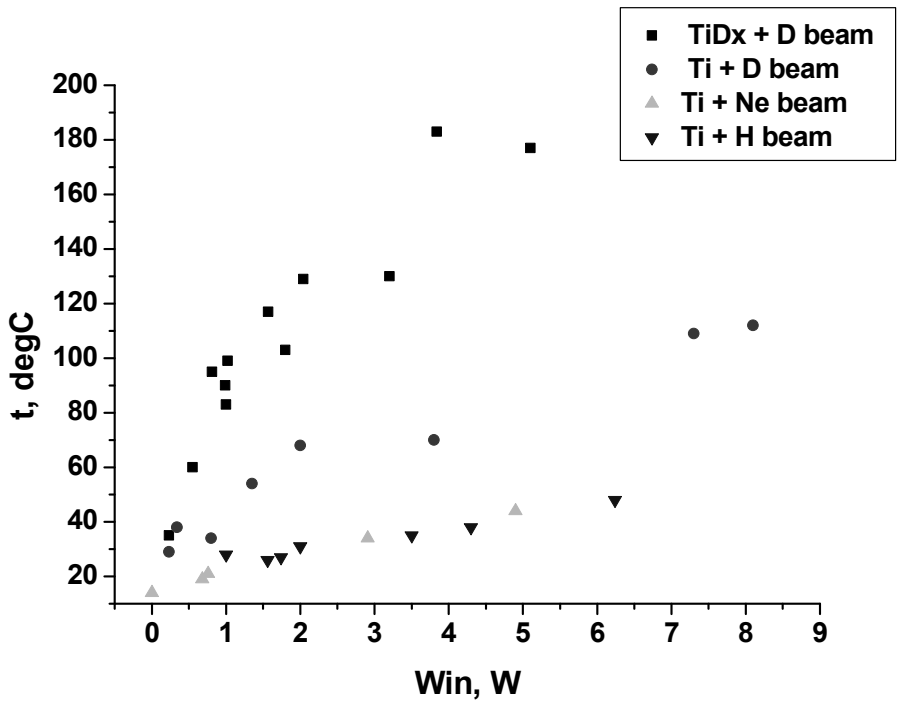

Fig.2. Dependence of the target temperature Ti/TiO$_2$:D$_x$ and Ti on the incoming power. ■ the Ti/ TiO$_2$: D$_x$ target 55 microns, the D+ beam. ● the Ti target, beam D+. ▲ the Ti target, the Ne+ beam. ▼ the Ti target Ti, the H+ beam.

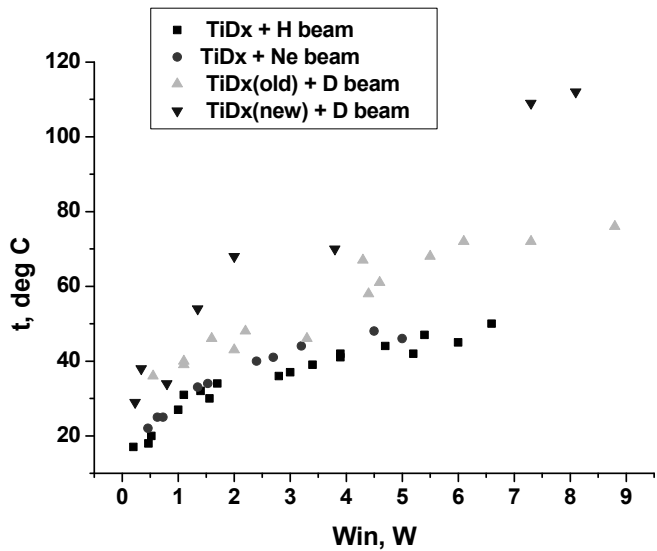

Fig. 3. The target temperature for Ti/TiO$_2$:D$_x$ sequentially irradiated with beams of D+, H+, and Ne+ in the same experiment, shown as a function of the incoming power. ■ the H+ beam. ● the Ne+ beam. ▼ the target at the start of the irradiation beam D+. ▲ the target at the end of the irradiation beam D+.

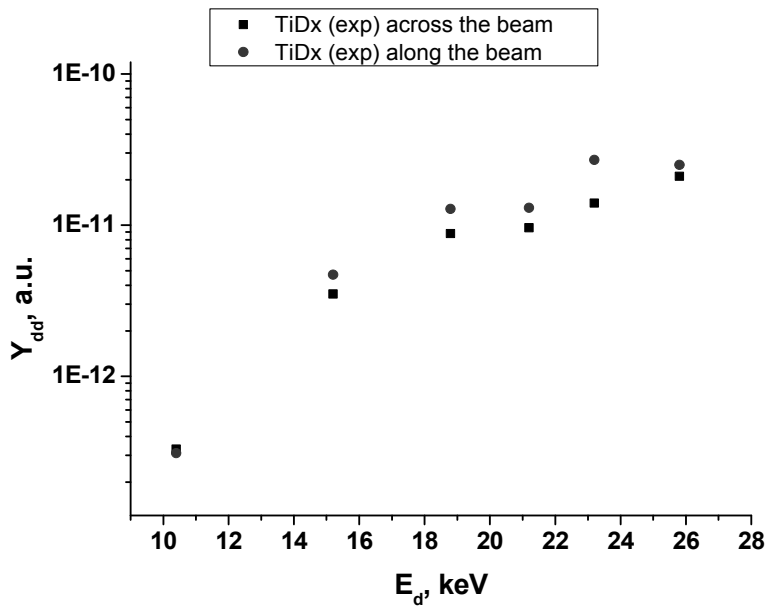

Fig. 4. Yield of the DD-reaction neutrons in deuterated Ti/TiO$_2$:D$_x$ target, irradiated by D+ beam shown as a function of the incoming energy. ■ yield of DD-reaction in the direction of the beam, ● across the direction of the beam.

The temperature dependence of the Ti/TiO$_2$:D$_x$ and Ti-targets on the incoming power, W$_{in}$, determined by the beam energy and current, is investigated. The targets were irradiated with the ion beams of D+, H+ and Ne+. The measurements demonstrated that the temperature of the deuterated targets of Ti/TiO$_2$:D$_x$ and Ti is much higher, when irradiating by a D+ beam, than the Ti-target temperature under irradiation with H+ and Ne+ beams, as illustrated in Fig. 2. An important fact is a non-linear dependence of the target temperature on the incoming power for Ti/TiO$_2$:D$_x$ target, exposed by the D+ beam, which is strongly affected by the current density of the deuterium-ion beam.

The temperature of Ti/TiO$_2$:D$_x$ target, continuously exposed with the beams of D+, H+ and Ne+ in one experiment, in dependence on the incoming power, W$_{in}$, is shown in Fig. 3. A gradual reduction of the target temperature is observed with decreasing deuterium concentration through thermal desorption. The temperature of the deuterated Ti/TiO$_2$:D$_x$ target is much higher in the irradiation with the D+ beam, as compared to the H+ and Ne+ beams.

A possible explanation of the heat-emission excess from the Ti/TiO$_2$:D$_x$ target irradiated with D+ beam, as compared to the Ti target irradiated with H+ or Ne+ beam can be derived from the assumption that in contrast to a DD-reaction in gas, such reaction in a solid-state target is possible in four channels:

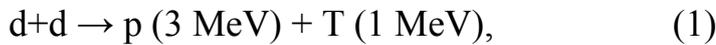

d+d → p (3 MeV) + T (1 MeV),        (1)

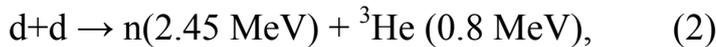

d+d → n(2.45 MeV) + $^3$He (0.8 MeV),        (2)

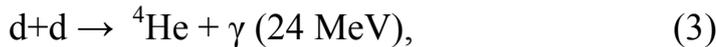

d+d → $^4$He + γ (24 MeV),        (3)

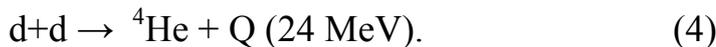

d+d → $^4$He + Q (24 MeV).        (4)

The reaction (4) would be the main channel, while reactions (1-3) are strongly suppressed at low energies [6]. This hypothesis is partially confirmed by our experimental measurements of the DD-reaction yield in the channel (2) along and across beam direction using $^3$He detector, as shown in Fig. 4.

The neutron yield in the energy range of 10 - 26 keV per beam-current unit does not exceed 10$^{-10}$ which can not explain the observed excess of heat emission from the Ti/TiO$_2$:D$_x$ target by «classical» channels of DD-reactions (1) and (2).

In the reaction (4), the main process responsible for the energy radiation of the compound nucleus $^4$He* to the ground state $^4$He is the exchange of virtual photons between the compound nucleus and the nearest electrons, including electrons of the crystal lattice. At the same time, the decays with nucleon emission are energetically forbidden.

In the reaction (4), the energy of about 24 MeV is released, which is utilized for heating up the whole crystal structure of the target. For attaining of power of 1 W, $2.6 \times 10^{11}$ such reactions would be necessary. During our experiment, (approximately $10^4$ s) the amount of about $10^{15}$ of helium atoms is formed, which is insufficient for mass-spectrometry diagnostics. The heat-emission excess is the only indication of the reaction (4), registered in our experiment.

Further, more accurate calorimetric experiments are necessary to clarify the origin of the excess of the heat emission in the deuterated target.

Oleg Dmitrievich DALKAROV
Michail Aleksandrovich NEGODAEV
Aleksei Sergeevich RUSETSKII
**Investigation of heat release in the targets during irradiation by ion beams**